\documentstyle[11pt]{article}
\begin{document}
\centerline{\bf Quantum Cryptography Based on the}
\vskip 2mm
\centerline{\bf Time--Energy Uncertainty Relation.}
\vskip 3mm
\centerline{S.N.Molotkov and S.S.Nazin}
\vskip 3mm
\centerline{\sl\small 
Institute of Solid State Physics of the Russian Academy of Sciences,}
\centerline{\sl\small 
Chernogolovka, 142432 Russia}
\vskip 3mm
\begin{abstract}
A new cryptosystem based on the fundamental time--energy uncertainty
relation is proposed. Such a cryptosystem can be implemented 
with both correlated photon pairs and single photon states.
\end{abstract} 

Quantum cryptography is based on the following two features of 
quantum theory. First, the stochastic nature of the measurement 
results in quantum mechanics allows two distant observes to perform 
measurements of an appropriate physical quantity on a system 
prepared in some special states to get a pair of perfectly correlated 
random sequences of zeros and units which can be used as a key 
[1--9]. Second, any eavesdropper or intruder attempting to affect the 
key generation procedure can be detected because of the fact
that generally the system wavefunction before the 
measurement cannot be determined from the results of that measurement 
[10].

Proposed in the present paper is a new cryptosystem whose
security is based on the quantum mechanical time--energy uncertainty 
relation.

The cryptosystem is based on a joint measurement of the
correlated two-photon (biphoton) states of the radiation field,
although it will be seen that {\it a cryptosystem of that kind 
can also be implemented with single-photon states with different 
frequencies, which is very important from the point of view of its 
physical realization.}

A biphoton field is written as
\begin{equation}             
|\Psi\rangle=\exp{(-i\omega_0t)}\int g(\omega ) |1\rangle_{\textstyle 
\omega}|1\rangle_{\textstyle \omega_0-\omega} d\omega, 
\end{equation} 
where $|1\rangle_{\textstyle \omega}$ is the single-photon state
with frequency $\omega$, $\omega_0$ is a certain fixed frequency 
(which is the sum of frequencies of two photons occurring in the 
biphoton), and $g(\omega)$ is the function defining the biphoton 
spectrum width.  

The key generation procedure employs the photodetectors with
different registration bandwidth. Suppose that the two users A and 
B register the photons with the photodetectors having central frequencies
$\omega_{A,B}$ and bandwidths $\gamma_{A,B}$. Let an atom be prepared
in the excited state at the moment $t=0$ and the photons are detected
by the users at $t_{A}^0$ and $t_{B}^0$, respectively. The biphoton 
field correlation function for a joint measurement by the two users
depends on the relative delay of reduced registration moments
$T=t_B-t_A$, where $t_{A,B}= t_{A,B}^0-r_{A,B}/c$ 
($r_{A,B}$ is the distance between the source and the users, and
$c$ is the velocity of light in the medium) [11], and can be 
written in the following form:
\begin{equation}             
P(T)=(2\pi\gamma_A\gamma_B)^2
\frac{\textstyle \theta 
(T)\exp{(-2\gamma_B T)}+\theta(-T)\exp{(2\gamma_A T)}}
{\textstyle \Omega^2+(\gamma_A+\gamma_B)^2},
\end{equation}
\begin{displaymath}
\Omega=\omega_0-\omega_A-\omega_B.
\end{displaymath}
Here $\theta(T)$ is the step function. 

If the biphoton field is measured by two narrow-band detectors
$(\gamma_{A,B}\rightarrow 0)$, they can only fire simultaneously if
$\omega_A+\omega_B=\omega_0$, i.e., when the users employ the
detectors with two complementary frequencies: 
\begin{equation}             
P(T)\propto\delta (\omega_0-\omega_A-\omega_B)\equiv
\mbox{photon energies are correlated.} 
\end{equation}
When the radiation field defined by Eq.(1) is measured by two wide-band
detectors $(\gamma_{A,B}\rightarrow\infty)$, it follows from Eq.(2) that 
\begin{equation}             
P(T)\propto\delta (t_B-t_A)\equiv
\mbox{photon detection moments are correlated.} 
\end{equation}
When employing the narrow-band photodetectors,  
($\gamma\rightarrow 0$, Eq.(3)), the difference between photon registration
times by the two users $T$ can have any value in the interval 
$-\infty < T <\infty$. 

On the other hand, if the wide-band photodetectors are used
($\gamma\rightarrow\infty$, Eq.(4)) the difference between
the reduced photon registration times is very close to zero
(to within $\sim 1/\gamma\rightarrow 0$); however, in this case
no information on the energy (frequency) of the photons is gained.

The key generation protocol consists of the following steps:
\begin{itemize} \itemsep -3pt
\item[1)] The users select two fixed frequencies  
$\omega_1$ and $\omega_2$  such that  $\omega_1 + \omega_2 = \omega_0$. 
In each measurement users A and B randomly (and independently of 
each other) choose either wide-band or one of the two narrow-band
photodetectors with central frequencies $\omega_1$ and $\omega_2$. 
\item[2)] After a series of measurements is completed, all the
measurements where at least one of the detectors did not fire are discarded.
\item[3)] The users announce through a public channel which photodetectors
(wide- or narrow-band) they used in each measurement, but do not specify
which of the two narrow band detectors ($\omega_1$ or $\omega_2$)
was used in a particular measurement.
\item[4)] Since only the measurements where both photodetectors fired
are considered, the measurements where both users employed the
narrow-band photodetectors yield two perfectly correlated random sequences
which can be used as a key. For instance, if user A employed the detector 
with $\omega_1$, he knows that the other detector could only fire if user
B employed the detector with $\omega_2$ (this event is regarded as
corresponding to logical 1), and vice versa, if user A employed the detector 
with $\omega_2$, user B necessarily employed the detector centered 
about $\omega_1$ (logical 0).
\end{itemize}
To detect possible eavesdropping at the key generation stage, the users
can analyze the measurements where they both employed the wide-band 
detectors. In that case users A and B actually measure the relative 
delay time $t_B-t_A$ (the line length is assumed to be known).
If no eavesdropping took place, $T = t_B-t_A=0$ (to within
$1/\gamma_{A,B}\rightarrow 0$); on the contrary, any eavesdropping
attempt would result in $T\ne 0$. Indeed, to obtain information on the key
(i.e., to determine the photon frequency), the eavesdropper should use
a narrow-band photodetector with the bandwidth $\gamma_*\rightarrow 0$, 
since a wide-band detector cannot distinguish between the photons with
energies $\omega_1$ and $\omega_2$, which carry the information on the key.
Since the photodetector types are chosen by users A and B in a random way,
the situations where users A and B employed wide-band detectors
(i.e., they should have $T=0$) while the eavesdropper employed a
narrow-band detector (measured the photon energy with the accuracy
$\gamma_*\rightarrow 0$) will inevitably occur. However, according to the
quantum mechanical time--energy uncertainty relation
\begin{equation}        
\Delta E\Delta t\geq \hbar,  
\end{equation} 
such a measurement cannot be performed during the time interval
shorter than $\hbar/\gamma_*$. Therefore, users A and B will be able
to detect eavesdropping by a systematic deviation of the
difference between the reduced photon registration moments 
$T=t_B-t_A$  from zero.  

It should be noted that the situation with Eq.(5) is not as simple
as for uncertainty relations for other variables, such as
coordinate and momentum, or different spin components. In spite of 
a long history [12--18], both the validity of relation (5) and 
interpretation of the quantities $\Delta E$ and $\Delta t$ are
still discussed in the literature (see a review article [16]).
According to Ref.[16], the problem of validity of relation (5)  
reduces to choosing between the following two statements:
\begin{itemize} \itemsep -3pt
\item[1)] Scatter of the energy of a quantum system $\Delta E$
before and after the measurement satisfies the inequality (5), where
$\Delta t$ is the measurement procedure duration, i.e., 
arbitrarily fast measurements which do not disturb the state of a 
quantum systems {\em are impossible}.
\item[2)] It is possible to realize {\sl arbitrarily fast}
measurements which do not disturb the quantum system energy, and, therefore,
violate inequality (5).
\end{itemize}
Strictly speaking, there are specific examples of explicitly 
time-dependent Hamiltonians (see e.g., Refs.[17,18]) violating
inequality (5). Hence, the problem of validity of Eq.(5)
reduces to the possibility (or impossibility) of the physical 
realization of a particular Hamiltonian allowed by quantum mechanics
as a formal mathematical scheme [16]; it should be emphasized here
that so far no physical realization of the Hamiltonians formally 
violating Eq.(5) has been proposed. At the same time, it is obvious 
that it is natural to consider only the physically realizable
Hamiltonians [12--16]. Therefore, at present there is 
every ground to consider Eq.(5) as a fundamental law of Nature.
We shall adopt this orthodox point of view [12--16] and assume 
that {\em it is impossible to precisely measure the photon energy 
in an arbitrarily short time interval.}

As an illustration, consider the situation where user A employs a wide-band 
photodetector $(\gamma_A\rightarrow\infty)$, while the other (eavesdropper
which is now playing the role of user B) employs a narrow-band one
$(\gamma_B\rightarrow 0)$. In this case user B always registers a photon 
with a certain 
delay time which can take any positive value with equal probability.
To understand this result qualitatively, the process of photodetection can
be viewed in the following way. We shall assume that the photodetector
consists of an absorber (two-level ``atoms'') which is tunnel-coupled to a
macroscopic ``electrode''. The electron lifetime at the
quasistationary level after the absorption of a photon is determined
by the level width (this lifetime is actually the measurement duration:
on the average, it is the time taken by electron to leave the quantum 
system and reach the classical one (electrode)). In a wide-band detector
($\gamma_A\rightarrow\infty$) the level width is large, so that the
absorbed photon energy cannot be precisely determined. However, in that case 
the photodetection process is fast because of a rapid 
electron transition to the electrode. In a narrow-band detector
($\gamma_B\rightarrow 0$) the photon is only absorbed if its frequency 
coincides with the separation between the two levels. On the other hand,
the photodetection process is slow, and in the limit $\gamma_B=0$ the
electron oscillates between the two levels [19] without tunneling into
the electrode, so that the measurement (photodetection) is formally 
infinitely long (it is this feature of the photodetection process which 
results in the delay of the photon registration time by user B).

Thus, if Eq.(5) holds (statement 1), it is impossible to measure 
the photon frequency by a narrow-band detector in an arbitrarily 
short time interval (this would violate Eq.(5)), and this circumstance
can be used to detect eavesdropping.

It should be emphasized that {\em the inevitable time delay
between the measurements of users A and B in the described 
protocol originates from the fundamental relation (5) rather
than the technical complexities introduced by the users.}
By increasing the fraction of measurements employing the wide-band 
photodetectors, the probability of detection of the eavesdropper 
can be made arbitrarily close to unity.

Obviously, the less is the difference between $\omega_1$ and $\omega_2$, 
the worse is the situation for the eavesdropper;  
currently the difference $|\omega_1 - \omega_2| \sim 10^5$ s$^{-1}$ at
the frequency $\omega_1 \approx \omega_2 \sim 10^{15}$ s$^{-1}$ seems 
to be easily achievable (for the wavelength 
$\lambda_0=1.3$ $\mu$m, corresponding to the optical fiber transparency 
window, $\omega \sim 10^{15}$ s$^{-1}$). To distinguish between the
photons with frequencies $\omega_1$ and $\omega_2$, the eavesdropper 
should measure the photon frequency with the accuracy better then
$10^{5}$ c$^{-1}$ which would result in the delay of order 
$10^{-5}$ s (i.e., effective increase in the line length by about 3 km).
At the same time the difference between the reduced photon 
registration moments $T$ in the absence of eavesdropping can be
made rather small even with not too wide-band photodetectors.
For the detector bandwidth $\gamma \sim 10^9$ s$^{-1}$ (i.e.,
of the order of atomic level natural linewidth) 
$T \sim 1/\gamma \sim 10^{-9}$ s (i.e., in the context of the present
problem a photodetector with the bandwidth $\sim 10^9$ s$^{-1}$ can already 
be regarded as a wide-band one). 
 
We conclude with the following important remark. A similar scheme
can be constructed without the biphoton states if user A can send 
single photons with close energies $\hbar\omega_1$ and $\hbar\omega_2$, 
for which the moments of their emission is known with sufficiently 
high accuracy. For example, one can use the quantum dots where the
electron lifetime at the excited size-quantized level is of the order
of $10^{-9}$ s: if electron is excited to such a level by a short
(picosecond) pulse, the photon emission time into the line can be
regarded as known to within $10^{-9}$ s. Therefore, if two such photon
sources with frequencies separated by  $\sim 10^5$ s$^{-1}$ are available,
they can be used to construct a cryptosystem based on the time--energy
uncertainty relation. Such a scheme has advantages over the interferometric
schemes since it does not require regular adjustment of the long-base 
interferometer [7]. The described cryptosystem also allows to distribute
the key over a network of equivalent users (by transferring it through a 
chain of users) over the distances exceeding the signal decay length 
in the optical fiber in a way similar to that described in Ref.[20].

We are grateful to  S.V.Iordanskii, M.V.Lebedev, S.T.Pavlov, 
and I.I.Tar\-ta\-kov\-sky for fruitful discussions.
This work was supported by the Russian Fund for Fundamental Research 
(Grant No 96-02-19396).

\end{document}